\documentclass[prd,amsmath,amssymb,aps]{revtex4-1}

\usepackage{graphicx}

\newcommand\cc[1]{#1^{^{\kern-6pt \circ}}\kern2pt}

\def\pa{\partial}

\newcommand{\m}{\mu}
\newcommand{\n}{\nu}

\def\be{\begin{equation}}
\def\ee{\end{equation}}
\def\bea{\begin{eqnarray}}
\def\eea{\end{eqnarray}}
\def\ba{\begin{array}}
\def\ea{\end{array}}
\def\bi{\begin{itemize}}
\def\ei{\end{itemize}}

\newcommand{\beq}{\begin{equation}}
\newcommand{\eeq}{\end{equation}}
\newcommand{\beqn}{\begin{eqnarray}}
\newcommand{\eeqn}{\end{eqnarray}}
\newcommand{\bga}{\begin{align}}

\def\dalemb#1#2{{\vbox{\hrule height .#2pt
\hbox{\vrule width.#2pt height#1pt \kern#1pt
\vrule width.#2pt}
\hrule height.#2pt}}}

\begin{document}

\title{\LARGE Singleton deformation of higher-spin theory and the phase structure of the three-dimensional $O(N)$ vector model}
\author{\bf Robert G. Leigh,} 
\affiliation{Department of Physics, University of Illinois\\1110 W. Green Street, Urbana IL 61801, U.S.A.}
\author{\bf Anastasios C. Petkou}

\affiliation{Department of Physics, University of Crete\\71003 Heraklion, Greece}

\email{rgleigh@illinois.edu}
\email{petkou@physics.uoc.gr}

\date{\today}                                          


\begin{abstract}
We consider a singleton deformation of the AdS$_4$ higher-spin theory dual to the three-dimensional $O(N)$ vector model. The singleton couples to the higher-spin multiplet only through a marginal boundary interaction. We argue that the effect of such a deformation is to shift $N$ $\to$ $N+1$ in both sides of the holographic correspondance and we show how the gap equations of the three-dimensional $O(N)$ vector model arise from the higher-spin theory. The singleton deformation breaks higher-spin symmetry and gives rise to  the well-known $1/N$ anomalous dimensions of the boundary theory. 

\end{abstract}
\maketitle

\section{Introduction and Summary}
Three-dimensional models have long been used as simple laboratories for a number of interesting field theoretic properties. Among the simplest and best known examples are the bosonic and fermionic (or Gross-Neveu) $O(N)$ vector models which are renormalizable field theories in the context of the $1/N$ expansion.\footnote{For reviews see \cite{Zinn,Rosenstein}.} These models exhibit critical behavior described by nontrivial CFTs with perturbatively calculable anomalous dimensions for all operators in their spectrum (e.g. \cite{Ruhl1,Ruhl2,Ruhl3,Ruhl4,Ruhl5,Tassos1,Tassos2}).  They also exhibit simple patterns of symmetry breaking. In particular, the bosonic model exhibits the  $O(N)\rightarrow O(N-1)$ global symmetry breaking pattern and the corresponding Goldstone mechanism, while the fermionic model exhibits parity symmetry breaking.

The advent of AdS/CFT has revived the interest for these models and in particular for the bosonic one due to its conjectured holographic duality \cite{KP} with the simplest higher-spin gauge theory on AdS$_4$.\footnote{For  reviews of higher-spin theories see e.g. \cite{Vasiliev,Vasiliev1}.}   It was also suggested in \cite{LP1} that the fermionic $O(N)$  model also has a higher-spin dual, termed the B-model in \cite{SS1},   the latter being the truncation to a conformally coupled pseudoscalar of ${\cal N}=1$ SUSY higher-spin theory in AdS$_4$. 
A different in nature higher-spin holography in the context of AdS$_3$/CFT$_2$ has also attracted interest lately \cite{Matthias,Kraus}.

While important progress in understanding the conjectures \cite{KP,LP1,SS1}  has been made in a number of recent works \cite{Yin1,Yin2,Maldacena,Jevicki}, some crucial questions remain open. A primary question  is whether the symmetry breaking patterns of the boundary three-dimensional theories can be understood holographically. Progress here stumbles on the poor understanding of the bulk counterpart of the $O(N)$ boundary symmetry. For example, it is not clear yet whether there are non-perturbative objects in higher-spin gauge theory on AdS$_4$ that give rise to the global $O(N)$ symmetry in the boundary. Hence, it is not known if only the singlet sector\footnote{For some recent work on gauging the boundary symmetries using Chern-Simons fields see \cite{Aharony,Minwalla}.} of global $O(N)$ can be described holographically, despite the fact that boundary vector theories possess nontrivial and well-understood non-singlet sectors as well \cite{Tassos1}.

In this note we focus on the bosonic $O(N)$ vector model and ask how higher-spin gauge theories in AdS$_4$ could describe holographically  its $O(N)\rightarrow O(N-1)$ symmetry breaking pattern. 
We show that this can be done by deforming the bulk theory by a scalar singleton ---  a  bulk scalar with mass $m^2=-5/4$ --- which is coupled to the higher-spin multiplet only via marginal boundary interactions. We will argue that if the initial bulk higher-spin theory was describing the $O(N)$ model, then the deformed theory describes the $O(N+1)$ model. As crucial evidence for our suggestion, we show that using our bulk construction we exactly reproduce the gap equations of the $O(N)$ theory that determine its vacuum structure.    
When extended by the singleton, the bulk theory breaks higher-spin symmetry due to the presence of the marginal boundary deformation. Nevertheless, the deformation leads to a diagrammatic $1/N$  expansion for the two-point functions of the boundary scalar leading to an $O(1/N)$ anomalous dimension. The latter turns out to coincide with the UV anomalous dimension of the $N$ elementary scalars in the $O(N)$ vector model. 

Our bulk construction indicates that singletons play a crucial role in the holography of vector-like theories.  Although it is often said that singletons decouple because they are dual (in alternative quantization) to free scalar fields, here they are made non-trivial through the imposed  marginal boundary interactions. Also, our results provide positive evidence for the point of view that the reduction to the singlet sector is a consistent truncation of the higher-spin theory and does not require gauging the boundary global symmetry. Finally,  the extension of our mechanism to include the boundary marginal interactions of singletons to the full higher-spin multiplet should be the analog of the introduction of a single $D3$-brane in five-dimensional supergravity such as to describe the boundary $SU(N)\rightarrow SU(N+1)$ symmetry enhancement. It would be very interesting to understand the singleton physics on a similar footing. 

In Section 2 we review the field theoretic description of the $O(N)$ vector model, its large-$N$ expansion, gap equations and $1/N$ anomalous dimensions of its low lying scalars. This material is standard, but we present it in such a way that comparison to the holographic construction is straightforward. In Section 3 we extend the bulk theory with bosonic singletons and use the appropriate boundary conditions to describe the boundary gap equations that describe clearly the $O(N)\rightarrow O(N+1)$ symmetry enhancement pattern in the boundary. Moreover, we remark that the presence of the boundary marginal deformation yields the  $1/N$ anomalous dimension of the elementary scalar field at the UV fixed point. In Section 4, we summarize and discuss the extension of our results to fermionic theories and general higher-spin holography.

\section{Review of the $O(N)$ vector model}
\subsection{The model and its large-$N$ expansion}
For simplicity we work in Euclidean space. The model consists of $N$ elementary scalar fields $\phi^a(x)$, $a=1,2,..,N$ with Lagrangian 
\beq
\label{ON}
L=\frac{1}{2}\int \!d^3x \,\partial_\m\phi^a\partial_\m\phi^a\,,
\eeq
subject to the constraint\footnote{Another model with a large $N$ critical point can be obtained by introducing $\phi^4$ interactions in the free theory. In that case, the $\rho$ field is obtained via the Hubbard-Stratanovich mechanism, and the resulting action differs from (\ref{I1}) in the form of the terms depending only on $\rho$. 
}
\beq
\label{ONconstr}
\phi^a\phi^a=\frac{1}{G}\,.
\eeq
Using a delta-function representation the constraint can be inserted into the partition function introducing a 
Lagrange multiplier scalar field $\rho$ as\footnote{The $\rho$ integration runs parallel to the imaginary axis.}
\bea
\label{Z1}
Z&=&\int ({\cal D}\phi^a)({\cal D}\rho)e^{-I(\phi^a,\rho)}\,,\\
\label{I1}
I(\phi^a,\rho)&=&\frac{1}{2}\!\int d^3x\,\phi^a(-\partial^2)\phi^a+\frac{1}{2}\int \!d^3x \,\rho\left(\phi^a\phi^a-\frac{N}{g}\right)\,.
\eea
We have  defined the rescaled coupling $g=GN$ anticipating a large-$N$ expansion. The dimensionfull coupling $1/G$ sets the physical mass scale of the theory. From (\ref{ONconstr}) we note that $G\rightarrow 0$ is the free field theory limit which lies in the UV.

Integrating out the $\phi^a$'s one obtains the effective action for the $\rho$ field,  $S_{eff}(\rho)$, as
\beq
\label{SeffON}
Z=\int({\cal D}\rho)e^{-NS_{eff}(\rho)}\,,\,\,
S_{eff}(\rho)=\frac{1}{2}{\rm Tr}\ln(-\partial^2+\rho)-\int d^3x\frac{\rho}{2g}\,,
\eeq
This integral has a saddle point at large-$N$ with constant $\rho_0$ determined by the gap equation
\beq
\label{gapON}
\frac{\partial S_{eff}(\rho)}{\partial\rho}\Bigl|_{\rho_0}=0\Rightarrow \frac{1}{g}=\int\frac{d^3p}{(2\pi)^3}\frac{1}{p^2+\rho_0}\,.
\eeq
Clearly, inspection of (\ref{I1}) shows that a constant value for $\rho_0$ constitutes a mass for the elementary scalars $\phi^a$ as $\rho_0=m^2$. 

The large-$N$ expansion is obtained by systematically expanding around this point as
\beq
\label{sigma0ON}
\rho(x)=\rho_0+\frac{1}{\sqrt{N}}\sigma(x)\,,
\eeq
and one obtains an effective action ${\cal S}_{eff}^N(\sigma,\rho_0)$ for the real fluctuations $\sigma$ as
\bea
\label{Seffs}
S_{eff}(\rho)&=& V_{eff}(\rho_0,g)+\frac{1}{N}{\cal S}^{N}_{eff}(\sigma,\rho_0)\,,\\
V_{eff}(\rho_0,g) &=& \frac{1}{2}{\rm Tr}\ln(-\partial^2+\rho_0)-\frac{\rho_0}{2g}(Vol)_3\,,\\
\label{Sefflambda}
{\cal S}^{N}_{eff}(\sigma,\rho_0) &=& \frac{1}{2}\int \,\sigma(x)\Delta(x,y;\rho_0)\sigma(y)+\frac{1}{3!\sqrt{N}}\int \sigma(x)\sigma(y)\sigma(z)P(x,y,z;\rho_0) +..
\eea
where $(Vol)_3$ is the three dimensional volume and $U_{eff}(\rho_0,g)=V_{eff}(\rho_0,g)/(Vol)_3$ the effective potential of the theory.  Using (\ref{Sefflambda}) one constructs  a generating functional $W[\eta]$ for connected correlation functions of $\sigma$ as
\beq
\label{Weta}
e^{W[\eta]}\equiv\int ({\cal D}\sigma)e^{-{\cal S}_{eff}^N(\sigma,\rho_0)+\int\eta\sigma}\,.
\eeq

\subsection{The $O(N)\rightarrow O(N-1)$ symmetry breaking}
The  gap equation (\ref{gapON}) determines the vacuum structure of the theory.  
Introducing a UV cutoff $\Lambda$ for the momentum integral this can be rewritten as
\bea
\frac{1}{g}&=& \int^\Lambda\!\!\frac{d^3p}{(2\pi)^3}\frac{1}{p^2}-\int^\Lambda\!\!\frac{d^3p}{(2\pi)^3}\frac{\rho_0}{p^2(p^2+\rho_0)}\nonumber\\
\label{gapexpl}
&=& \frac{\Lambda}{2\pi^2}-\frac{\sqrt{|\rho_0|}}{2\pi^2}\arctan\frac{\Lambda}{\sqrt{|\rho_0|}}\,.
\eea
 If we define a critical coupling  $g_*$ as
\beq
\label{g*}
\frac{1}{g_*}=\frac{\Lambda}{2\pi^2}\,,
\eeq
then (\ref{gapexpl}) takes the suggestive form
\beq
\label{gapON2}
\left(\frac{1}{g_*}-\frac{1}{g}\right)=\frac{\sqrt{|\rho_0|}}{2\pi^2}\arctan\frac{\Lambda}{\sqrt{|\rho_0|}}=\frac{\sqrt{|\rho_0|}}{4\pi}+O(\rho_0/\Lambda)\,,
\eeq
We see that the vacuum structure depends essentially on the value of the bare coupling $g$ with respect to the critical coupling $g_*$. The gap equation, as we have written it here, has a solution when $g>g_*$ in which the theory is massive, $m=\sqrt{|\rho_0|}\neq 0$. When we tune to $g=g_*$, there is no mass scale in the theory and (\ref{Sefflambda}) yields the generating functional of connected correlation functions of a scalar operator $\sigma$ with dimension $\Delta=2+O(1/N)$ in a three-dimensional CFT, as we will discuss in Sec. 2.3.  The latter is what is often called the {\it critical} $O(N)$ vector model. 

For $g<g_*$ the only solution of the gap equation is at $\rho_0=0$, however an arbitrary mass scale remains in the theory even after sending the cutoff to infinity. This is an indication that the theory enters a symmetry broken phase. The clearer way to see this is to separate  out the $N$'th component of $\phi^a$'s, which we denote as $\phi$, and integrate over the remaining $N-1$ elementary scalars to obtain
\beq
\label{Z2}
Z=\int [{\cal D}\phi][{\cal D}\rho]\ e^{-(N-1)S_{eff}(\rho,\phi)}\,,
\eeq
with the effective action now defined as
\bea
\label{SeffN-1}
S_{eff}(\phi,\rho) &=& S_{eff}^{N-1}(\rho)+\frac{1}{2(N-1)}\int \!d^3x \,\phi(-\partial^2 +\rho)\phi\,, \\
\label{Seff1}
S_{eff}^{N-1}(\rho)&=& \frac{1}{2}{\rm Tr}\ln(-\partial^2+\rho)-\frac{N}{(N-1)}\int \!d^3x\,\frac{\rho}{2g}\,.
\eea
Apart from the different $N$ scaling of the coupling constant $g$, the effective action $S_{eff}^{N-1}(\rho)$ is essentially the same as $S_{eff}(\rho)$ given in (\ref{SeffON}). 
The large-$N$ expansion is now performed around the constant saddle points $\rho_0$ and $\phi_0$ defined as
\beq
\rho(x)=\rho_0+\frac{1}{\sqrt{N-1}}\sigma(x)\,,\,\,\phi(x)=\phi_0+\varphi(x)\,.
\eeq
with $\rho_0, \phi_0$ determined by the  gap equations
\bea
\label{gap1}
\frac{\partial S_{eff}}{\partial \rho}\Bigl|_{(\phi_0,\rho_0)}=0 &\Rightarrow & \frac{\phi^2_0}{N-1} =\frac{N}{(N-1)}\frac{1}{g}-\int\!\frac{d^3p}{(2\pi)^3}\frac{1}{p^2+\rho_0}\,, \\
\label{gap2}
\frac{\partial S_{eff}}{\partial \phi}\Bigl|_{(\phi_0,\rho_0)}=0&\Rightarrow & \rho_0\phi_0=0\,.
\eea
The effective action is then written as
\bea
\label{Seffsphi}
S_{eff}(\phi,\rho) &=& V_{eff}(\phi_0,\rho_0) +\frac{1}{N-1}{\cal S}_{eff}^{N-1}(\varphi,\sigma)\,,\\
V_{eff}(\phi,\rho_0) &=& \frac{1}{2}{\rm Tr}\ln(-\partial^2+\rho_0)-\frac{N}{2(N-1)g}\rho_0(Vol)_3+\frac{1}{2(N-1)}\rho_0\phi_0^2(Vol)_3\,,\\
\label{Seff}
{\cal S}_{eff}^{N-1}(\varphi,\sigma) &=& {\cal S}^{N-1}_{eff}(\sigma,\rho_0) +\frac{1}{2}\int \,\varphi(x) D_0(x,y;\rho_0)\varphi(y)\nonumber \\
&&+\frac{1}{2\sqrt{N-1}}\int \sigma(x)\varphi^2(x)+\frac{\phi_0}{\sqrt{N-1}}\int\sigma(x)\varphi(x)\,.
\eea
Comparing (\ref{Seffs}) and (\ref{Seffsphi}) 
we see that the effective action for the  $O(N)$ model can be obtained if we take the effective action ${\cal S}_{eff}^{N-1}(\sigma,\rho_0)$ of the  $O(N-1)$ model and {\it integrate in} an elementary scalar field $\varphi$ with a marginal deformation $\int\sigma\varphi^2$, as well as a linear interaction $\int\varphi\sigma$ whose strength is proportional to the constant saddle point value $\phi_0$. In particular, at the critical point where $\rho_0=\phi_0=0$, we learn that the effective action of the {\it critical} $O(N)$ model is obtained from the effective action of the {\it critical} $O(N-1)$ model by integrating in a massless elementary scalar $\varphi(x)$ with marginal  interaction. Namely, the $O(N-1)$ model ``eats" elementary scalars with $O(1/\sqrt{N})$ marginal interactions by enlarging its symmetry, shifting $N-1\rightarrow N$. This is the observation that will allow us to describe holographically the $O(N)\rightarrow O(N-1)$ symmetry breaking.


Now let us return to the gap equations (\ref{gap1}--\ref{gap2}). As before, we set $\rho_0=m^2$ with $m$ the common mass of the fundamental fields $\phi^a$. Using then (\ref{gapexpl}) we obtain 
\beq
\label{gap11}
\frac{\phi_0^2}{N-1}=\left(\frac{N}{N-1}\frac{1}{g}-\frac{1}{g_*}\right)+\frac{|m|}{4\pi}+\cdots\,,
\eeq
Equation (\ref{gap11}) differs from (\ref{gapON2}) in two ways. Firstly, we notice the presence of an extra term on the left-hand side. Secondly, there is an extra $N/(N-1)$ factor in front of the coupling constant $1/g$. These two differences are intimately related. 

The modified gap equation (\ref{gap11}) yields an explicit manifestation of the Goldstone mechanism. 
From (\ref{gap2}) we see that $\phi_0$ and the physical mass $|m|$ cannot be nonzero simultaneously. Moreover, the physical mass must satisfy $|m|<\Lambda$.  We see that the free field theory regime lies in the UV since there $g<Ng_*/(N-1)$ and $g$ has dimensions of inverse mass. In this regime the mass  vanishes but from (\ref{gap11}) we see that we always have $\phi_0\neq 0$ unless we fine tune the theory exactly to the critical coupling $g=Ng_*/(N-1)$. Hence, once we move away from the UV free field theory, the $O(N)$ symmetry is always broken to $O(N-1)$ as one of the components of the elementary fields $\phi^a$ acquires a nonzero expectation value. As usual we also have $N-1$  Goldstone bosons which are seen here as the massless elementary scalars that were integrated out.  

When the coupling is tuned to $g=Ng_*/(N-1)$ we have $\phi_0=m=0$ and we arrive at the critical $O(N)$ vector model. Notice, however, that the above critical point differs from the critical point found in (\ref{gapON2}) which required tuning the bare coupling constant exactly to $g=g_*$. However, by writing as
\beq
\label{newfp}
\frac{N}{N-1}\frac{1}{g}-\frac{1}{g_*}=\frac{1}{g}-\frac{1}{g_*}+\frac{1}{N-1}\frac{1}{g}\,,
\eeq
we see that the critical point determined by (\ref{gap11}) is shifted away from being exactly $1/g_*$ by a quantity of order $1/(N-1)$ whose effect is to renormalize to zero the square of the condensate $\phi_0^2$ in (\ref{gap11}). This way we understand the differences of (\ref{gap11}) and (\ref{gapON2}) and we see how both of them lead to the same nontrivial critical theory in the IR. It is interesting to notice that we have reached this IR theory through a path where the $O(N)$ symmetry is always broken except at the two end points. 

Finally, as the coupling increases to $g>Ng_*/(N-1)$  the only way to satisfy (\ref{gap11}) is to have $\phi_0=0$,  but then we must also have $m\neq 0$. This means that the  $O(N)$ theory enters its massive phase. The common mass of the $N$ elementary fields is 
\beq
m=\frac{2\Lambda}{\pi}\left(1-\frac{N}{N-1}\frac{g_*}{g}\right)\,,
\eeq
and is smaller than the cutoff as should be required. It is only in the limit $g\rightarrow\infty$ that the physical mass approaches the cutoff.

We can now return to (\ref{Seffsphi}) and clarify further the relationship between the  $O(N)$ and $O(N-1)$ models. We note that the value of the critical coupling $g_*$ is independent of $N$. Starting then from an $O(N-1)$ model, the absorption of the elementary scalar $\phi$ is done once we enter the {\it massive} phase of the theory, namely when $g=Ng_*/(N-1)>g_*$. Then  it is possible to deform the theory by a marginal coupling and return to the universal fixed point at $g_*$, having however enlarged the symmetry to $O(N)$. Starting deeper in the massive phase with $g>Ng_*/(N-1)$ the model absorbs the elementary scalar and flows to the massive phase of the $O(N)$ model under the marginal deformation. Finally, if we start from $g<Ng_*/(N-1)$, then the addition of the elementary scalar can be done only when $\phi_0\neq 0$ and the linear interaction term $\int\sigma\varphi$ is nontrivial.  In this phase we can shift the scalar fluctuation as
\beq
\varphi=\hat{\varphi}+\frac{\phi_0}{\sqrt{N-1}}\frac{1}{-\partial^2}\sigma\,,
\eeq
A short calculation then gives 
\beq
\label{ZZ1}
Z=e^{-(N-1)V_{eff}(0,\phi_0)}\int ({\cal D}\hat\varphi)({\cal D}\sigma)e^{-\left[{\cal S}^{N-1}_{eff}(\sigma,0)+\frac{1}{2}\int\hat\varphi D_0\hat\varphi +\frac{1}{2\sqrt{N-1}}\int\sigma\hat\varphi^2-\frac{\phi_0^2}{2(N-1)}\int\frac{1}{-\partial^2}\sigma^2+..\right]}\,.
\eeq
The last term in the exponent of (\ref{ZZ1}) is a nonlocal version of the irrelevant double-trace deformation $\int\sigma^2$ which drives the theory in the UV where we expect to find the free $O(N)$ model. 
Clearly, all the above discussion can be repeated if we shift $N\to N+k$, $k\in \mathbb{Z}$  in which case we are describing the generic symmetry breaking pattern $O(N+k)\to O(N+k-1)$. In Fig.1 we draw the phase diagram of the bosonic model and in  Section  3.1 we will see how this picture arises holographically.

\begin{figure}
\centering
\includegraphics[width=0.7\textwidth]{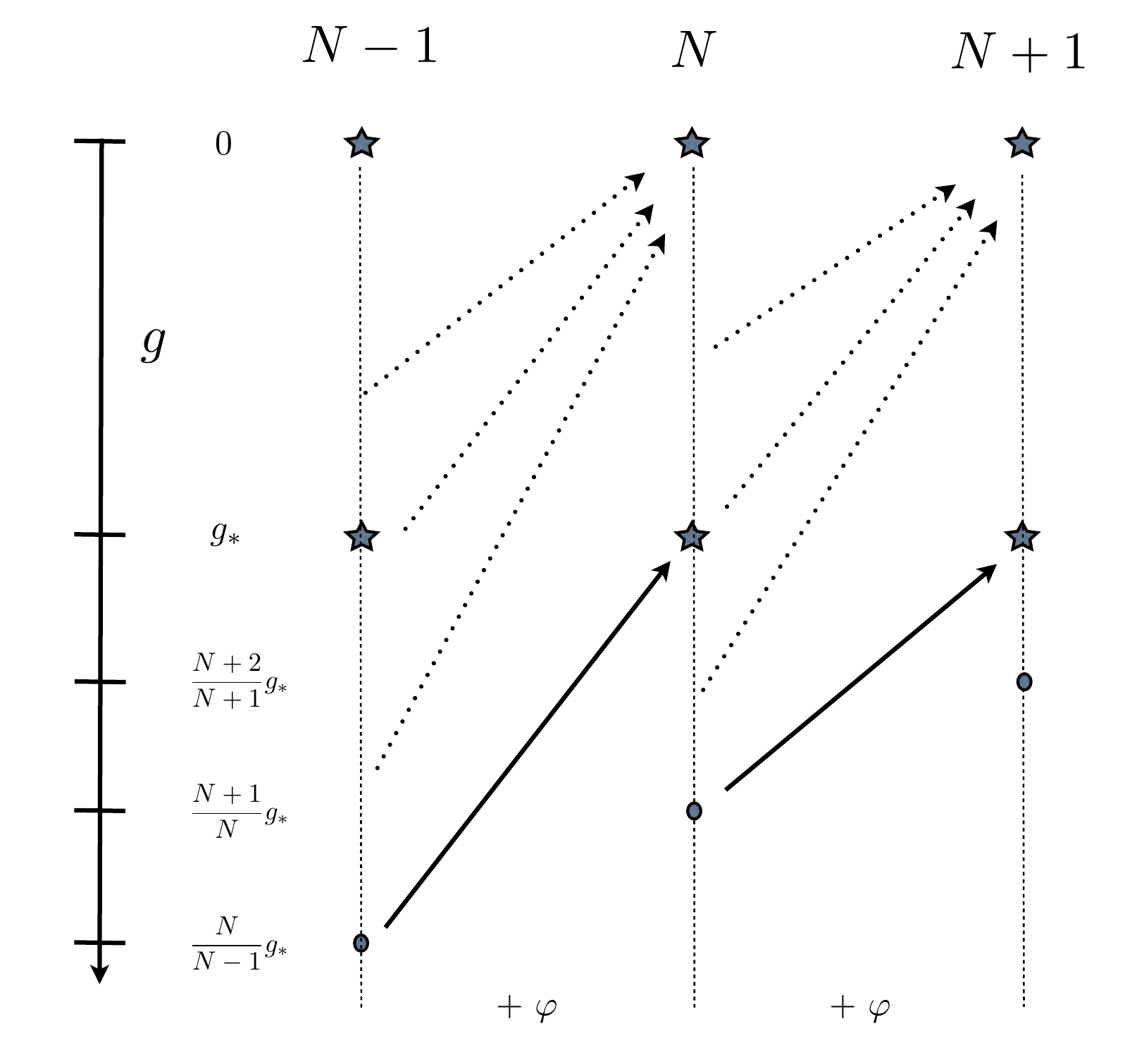}
\caption{The phase diagram of the bosonic $O(N-1), O(N)$ and $O(N+1)$ vector models and their relationships. The stars are the fixed points (CFTs). The solid arrows denote marginal deformations leading to the IR fixed point of the symmetry enhanced theory after the absorption of an elementary scalar $\varphi$. The dotted arrows denote deformations that include irrelevant double-trace ones leading to the UV fixed point of the symmetry enhanced theory. }
\label{Fig.1}
\end{figure}

%
%
%

\subsection{Anomalous dimensions in the critical $O(N)$ vector model}

The Lagrangian formulation of the $O(N)$ model allows for the calculation of all correlation functions of the operators $\phi^a$ and $\sigma$. For that, we need the partition function coupled to sources for $\phi^a$ and $\rho$ as
\beq
\label{Zphisigma}
Z\rightarrow Z[J^a,\eta]=\int[{\cal D}\phi^a][{\cal D}\rho]\ e^{-I(\phi^a,\rho)+\int \phi^a J^a+\int \eta\rho}\,.
\eeq
At $g=g_*$ this gives the generating functional\footnote{We have  appropriately rescaled the sources $J^a$ and $\eta$ in (\ref{WON}) such that the two-point functions for $\phi^a$ and $\lambda$ are $O(1)$.} for the critical $O(N)$ model as
\beq
\label{WON}
Z[J^a,\eta]=e^{W[J^a,\eta;g_*]}=e^{-NV_{eff}(0,g_*)}\int [{\cal D}\sigma]\ e^{-S_{eff}^{N}(\sigma,0)+\int \eta\sigma+\frac{1}{2}\int J^aD_0(\frac{i}{\sqrt{N}}\sigma)J^a}\,.
\eeq

Using (\ref{WON}) one can perform a systematic $1/N$ expansion for all correlation functions of $\phi^a$ and $\sigma$. However, the straightforward approach is not very useful beyond the leading order in $1/N$ since the  massless loop integrals that appear are unmanageable.  To deal with this issue, sophisticated methods that exploit the conformal invariance of the model have been devised in \cite{Khonkonen1,Khonkonen2,Khonkonen3} and have led to the calculation of the anomalous dimensions of $\phi^a$ and $\sigma$ to $O(1/N^3)$. These methods have also been applied  to fermionic  \cite{Gracey1,Gracey2,Gracey3} and supersymmetric \cite{Gracey4,Gracey5} vector models. 

At the same time,  W. R\"{u}hl and collaborators in a series of works (e.g. \cite{Ruhl1,Ruhl2,Ruhl3}) initiated the study of the operator spectrum and anomalous dimensions of the critical $O(N)$ vector model. Their methods have been extended by \cite{Tassos1,Tassos2,Tassos3} where it was shown that there exists a non-Lagrangian (bootstrap)  CFT formulation of the model where the basic dynamical information is the cancellation of shadow singularities in the OPE.  In this approach, all information regarding the operator spectrum comes from the OPE analysis of four-point correlation functions and one explicitly demonstrates the presence of an infinite series of higher-spin currents which are conserved in the large-$N$ limit, but acquire $1/N$ anomalous dimensions given by
\beq
\label{HSanomdims}
\Delta_s = s+1 +4\gamma_\phi\frac{s-2}{2s-1}\,,\,\,\,\,s=2k\,,\,\,\,\,\,k=1,2,..\,,
\eeq
where $\gamma_\phi$ is the anomalous dimension of $\phi^a$ to be calculated below. Notice that since the   higher-spin currents of the $O(N)$ model are operators of the form
\beq\label{hscurrents}
J_{(s)}\sim \phi^a \pa_{\{\mu_1}....\pa_{\mu_s\}}\phi^a\,
\eeq
with total symmetrization and tracelessness imposed on space-time indices, formula (\ref{HSanomdims}) is a remarkable check of an old argument by Parisi \cite{CallanGross} regarding the anomalous dimension of leading twist operators in non-asymptotically free theories. Indeed, for large $s$, the partial derivatives completely split the two elementary fields $\phi^a$ and the energy (i.e. scaling dimension) of a state described by (\ref{hscurrents}) tends to the sum of energies (i.e. dimensions) of the two elementary ``partons" $\phi^a$. This is in sharp contrast with the behavior of higher-spin currents in asymptotically free theories where the presence of gauge fields leads to the $\ln s$ scaling of anomalous dimensions for large $s$.  

Finally, let us briefly review the leading order calculation of anomalous dimensions for the elementary fields $\phi^a$ and $\sigma$. Using (\ref{WON}) it is straightforward to obtain the graphical $1/N$ expansion for the two-point function $\langle\phi^a\phi^b\rangle $ and $\langle\sigma\sigma\rangle$. 
Conformal invariance dictates their form to be
\beq
\label{ppll}
\langle\phi^a(x)\phi^b(0)\rangle=\frac{C_\phi}{x^{2\Delta_\phi}}\delta^{ab}\,,\,\,\,\,\langle\sigma(x)\sigma(0)\rangle=\frac{C_\sigma}{x^{2\Delta_\sigma}}\,.
\eeq
One way to do the calculations is to fix the dimension $d=3$ and define three critical indices $\gamma_\phi$, $\kappa$ and $z$ as
 \beq
\label{gphigl}
\Delta_\phi = \frac{1}{2}+\gamma_\phi\,,\,\,\,\Delta_\sigma=2-2\gamma_\phi-2\kappa\,,\,\,\,\,C^2_\phi C_\sigma=\frac{1}{\pi^4}+z
\eeq
with $\gamma_\phi,\kappa,z\sim O(1/N)$. 
As an example, consider the two-point function of $\phi^a$ which to leading order is given by $\sigma$-exchange. The loop integral can be performed using (\ref{integral}) and one finds
\bea
\label{pp1/N}
\langle\phi^a(x)\phi^b(0)\rangle &=& \frac{1}{4\pi}\frac{1}{|x|}\left[1+\frac{1}{N}\frac{4}{3\pi^2}\Gamma(-1+2\kappa)\frac{1}{|x|^{-4\kappa}}+...\right]\delta^{ab}\,,\nonumber\\
&=&   \frac{1}{4\pi}\frac{1}{|x|}\left[1-\frac{1}{N}\frac{4}{3\pi^2}\ln |x|^2+...\right]\delta^{ab}\,.
\eea
The logarithmic term in the rhs of (\ref{pp1/N}) gives the well-known result for the anomalous dimension of $\phi^a$ as\footnote{See for example the book \cite{Zinn_book} where the result quoted in Sec. 28 corresponds actually to $2\gamma_\phi$.}
\beq
\label{gphi}
\gamma_\phi=\frac{4}{3\pi^2}\frac{1}{N}
\eeq
For the calculations of $\kappa$ and $\zeta$ one needs to consider the 2-pt function of $\sigma$ and also the renormalisation of the vertex $\sigma\phi^2$. The most updated results are already a few decades old e.g. see \cite{Khonkonen3}. 

\section{The holographic description of the $O(N)$ vector model via higher-spin theory on AdS$_4$}

It was proposed in \cite{KP} that the free bosonic $O(N)$ model has a holographic dual on AdS$_4$ with higher-spin symmetry. The basic observation is that the free currents (\ref{hscurrents}) correspond to unitary irreducible representations (UIR) of $SO(3,2)$, denoted as $D(\Delta,s)$ with  dimensions $\Delta=s+1$. When $s$ is even, these arise in the parity-even tensor product of two singleton UIRs $D(1/2,0)$ as
\beq
[D(1/2,0)\otimes D(1/2,0)]_S=D(1,0)\oplus \sum_{s=1}^\infty D(2s+1,2s)\,.
\eeq
in which the lowest dimension operator (i.e. the "spin-zero" current) is a scalar of dimension $\Delta=1$. The suggested correspondence would proceed by considering a bulk action for the higher-spin gauge fields with the schematic form\footnote{The equations of motion of a consistent interacting higher-spin theory can be formulated on AdS$_4$ and have been written down by Vasiliev (see e.g. \cite{Vasiliev}). They necessarily include an infinite number of higher-spin fields. The Lagrangian formulation of the full set of equations is not yet known.}
\beq
\label{HSaction}
I_{HS}=\sum_{s=0,2,4,..}^{\infty}\int d^4 x\sqrt{-g}\frac{1}{2}\Phi^{(s)}\left[\Box_{s}-\frac{1}{L^2}(s^2-2s-2)\right]\Phi^{(s)}+O(1/\sqrt{N})\,.
\eeq
In (\ref{HSaction}), $\Phi^{(s)}$ denote symmetrized and double-traceless rank-$s$ tensors, $\Box_{s}$ are generalized Pauli-Fierz operators on the fixed AdS$_4$ background metric $g_{\m\n}$, and $(s^2-2s-2)/L^2$ is a mass term that is necessary to maintain higher-spin gauge invariance on AdS$_4$. By the usual rules of AdS/CFT, the quadratic part of the bulk action (\ref{HSaction}) yields the two-point functions of all free higher-spin currents (\ref{hscurrents}) normalized to $O(1)$. More precisely, since $\Phi^{(0)}$ is a conformally coupled scalar, in order to obtain the two-point function of $D(1,0)$ in the boundary one needs to quantize using the so-called {\it alternative quantization} AQ. The cubic interaction terms in  (\ref{HSaction}) would then give rise to the three-point functions of the $O(N)$ model which scale as $1/\sqrt{N}$. Higher order interaction terms would give rise to higher-point correlation functions in the boundary. 

Upon introduction of interactions, the free $O(N)$ theory flows down  to the IR critical point in which a dimension $\Delta=2$ operator, namely the UIR $D(2,0)$, is present in the spectrum. There, higher-spin symmetry is broken since the currents (\ref{hscurrents}) acquire nonzero anomalous dimensions of order $1/N$. Nevertheless, higher-spin symmetry is restored at least at $N\to\infty$. In Ref. \cite{KP}, this mechanism was demonstrated holographically by introducing the `double-trace' deformation $(\phi^a\phi^a)^2$ to the bulk action and noting that this has the same effect as the Legendre transformation that switches the quantizations of the bulk conformally coupled scalar field.

We will review this mechanism below from a new point of view. In particular, it is our intent to recover the detailed form of the boundary gap equations, and thus the vacuum structure of the $O(N)$ vector model, from a holographic analysis. As we shall see, this requires a detailed examination of boundary terms in the on-shell action. Furthermore, it will be convenient, for reasons that we discuss below, to extend the bulk theory by the introduction of an additional scalar field, that may be thought of in terms of the dual field theory as the introduction of the Lagrange (or Hubbard-Stratanovich) field $\sigma$ that played a central role in the analysis of the last section. This apparently violates the higher spin symmetry, but we will find that that symmetry is restored in precisely the way expected from the field theory analysis. Finally, the understanding of the symmetry breaking phase structure will require the introduction of an additional bulk field, that is a singleton (i.e., in the UIR $D(1/2,0)$). We break the analysis into these two pieces to most closely follow the field theory analysis that we reviewed above.

\subsection{The gap equations from holography}

The output of any holographic calculation is the generating functional of correlation functions, $W[J]$, in which the source $J$ for an operator in the dual field theory is obtained as one of the asymptotics of a bulk field, the other asymptotic corresponding to the vacuum expectation value of the operator. The generating functional is obtained directly from the bulk action, evaluated on-shell in terms of the asymptotics of the bulk fields. This on-shell action is in general supplemented by boundary terms that renormalize the theory as well as by finite boundary terms that modify boundary conditions. As usual in field theory, the generating functional may be Legendre transformed to give a quantum effective action $\Gamma[\langle{\cal O}\rangle]$ depending on the one-point function of the corresponding operator. It is this object whose critical points determine the vacuum structure of the theory. As we review in the appendix, a Lagrangian deformation of the classical field theory action by the functional $f({\cal O})$ of the operator ${\cal O}$, corresponds at least at large $N$ to a simple deformation of the quantum effective action
\beq
\Gamma_f[\sigma]=\Gamma_0[\sigma]+f(\sigma)\,,\,\,\,\,\sigma=\langle{\cal O}\rangle\,.
\eeq
Thus, given such a deformation, the gap equation will be obtained as
\beq
\frac{\delta \Gamma_f}{\delta \sigma}\Big|_{\sigma=\sigma_*}=0
\eeq
where $\sigma_*$ denotes a solution to the gap equations. We see that the boundary Lagrangian deformation, denoted here by $f(\sigma)$, will directly impact the form of the gap equations. 
Given such a deformation, the induced change in the generating functional will be generically rather complicated, except in the `double trace' case, where we take $f$ to be quadratic -- then the Legendre transform back to $W[J]$ is linear and easily performed. For higher order polynomials, it is non-linear and a `Maxwell construction' is generally required.

The higher spin theory action (\ref{HSaction}) on $AdS_4$ includes the  bulk scalar field $\Phi^{(0)}\equiv \Phi$  of mass $m^2L^2=-2$. Asymptotically, this field behaves as
\beq
\Phi\sim \alpha z+\beta z^2
\eeq
In this particular case, we have a choice: the standard quantization (SQ) assigns $\alpha$ as the source for a $\Delta=2$ operator with vev $\beta$. The alternate quantization (AQ) instead interprets $\beta$ as the source for a $\Delta =1$ operator with vev $\alpha$. It is the alternate quantization then that gives rise to the free UV fixed point, with its $\Delta=1$ scalar operator, $\phi^a\phi^a$.

To mimic the field theory analysis, we propose extending the bulk theory to contain two fields with $m^2L^2=-2$, namely
\beq
\label{HSactionext}
I_{extHS} = I_{HS} +\int d^4x\sqrt{-g}\frac{1}{2}\Sigma\left[\Box+\frac{2}{L^2}\right]\Sigma\,.
\eeq
We will take $\Phi$ in AQ, and $\Sigma$ in SQ. Asymptotically, we write
\beqn
\Phi &\sim& \alpha z+\beta z^2\,,\\
\Sigma&\sim& \eta z+\sigma z^2\,,
\eeqn
so that $\Phi$ gives rise to a $\Delta =1$ operator with vev $\alpha$, while $\Sigma$ gives rise to a $\Delta=2$ operator with vev $\sigma$. Further, we suppose that (at least at the present order of discussion), these fields do not mix in the bulk (or via their internal boundary conditions). This means that the regularity conditions of the bulk equations yield $\alpha=\alpha(\beta)$ and $\sigma=\sigma(\eta)$, and determine the boundary generating functional as
\beq
\label{Wbetaeta}
I_{extHS} \rightarrow W[\beta,\eta]=\int\alpha(\beta)\beta-\int\sigma(\eta)\eta\,.
\eeq
It is important to point out  the different relative signs in (\ref{Wbetaeta}) which arise because of the opposite quantizations used for the bulk fields.  In particular, the on-shell bulk action equals {\it minus} the boundary  generating functional if one uses SQ
and the generating functional will be $W[\beta,\eta]$. Also note that starting from (\ref{HSactionext}) the two-point functions of both the operators with $\Delta=1$ and $\Delta=2$ are normalized to $O(1)$. This means, for example, that in terms of the elementary fields $\alpha \sim (\phi^a\phi^a)/\sqrt{N}$.  

If this were the full story, then constructing $\Gamma[\alpha,\sigma]$ would give no sign of the gap equation of the $O(N)$ model, as $\Sigma$ is decoupled from $\Phi$ (as well as the rest of the higher spin fields). This can be remedied by a rather mild modification. Namely, we introduce boundary terms that couple the two fields together, that is, we introduce a Lagrangian deformation of the form\footnote{The cubic potential written here is in fact the most general, up to a constant shift in the definitions of $\sigma$ and $\alpha$. The $\phi^4$ model is obtained by taking instead $V(\sigma)\sim\sigma^2$.}
\beq
\label{fas}
f(\alpha,\sigma)=\int\left(\alpha\sigma+V(\sigma)-\frac13 \lambda(\alpha-h)^3\right)\,,\,\,\,V(\sigma)=-\frac{\lambda'}{g}\sigma\,.
\eeq
Up to an overall normalization the deformation (\ref{fas}) depends on the two dimensionless constants $\lambda$ and $\lambda'$, while $h$ is a parameter with dimensions of mass. 
 Then we have 
\beq
\label{Gas}
\Gamma[\alpha,\sigma]=\int\left(\frac12 \alpha K_1\alpha-\frac12\sigma K_1^{-1}\sigma+\sigma(\alpha-\frac{\lambda'}{g})-\frac13 \lambda(\alpha-h)^3\right)
\eeq
where $K_1$ is an appropriate kernel.\footnote{Notice that the different signs arising from the different quantizations ensures the positivity of the quadratic kernels (see Appendix A).}   For constant $\alpha$ and $\sigma$, we obtain the gap equations\footnote{Since we have not written a constant term in $\Gamma$ but do have a linear term in $\sigma$, we are free to shift $\sigma$ by a constant. Thus, the RHS of  eq. (23) can always be chosen to be a perfect square.}
\beqn
\label{ag}
\alpha&=&\frac{\lambda'}{g}\\
\label{sah}
\sigma&=&\lambda (\alpha-h)^2
\eeqn
 The first equation (\ref{ag}) above is what we expect for the 1-point function of the $\sigma$-model and corresponds to the constraint (\ref{ONconstr}). Taking into account the rescaling of the coupling which we have done after (\ref{I1}) we conclude that $\lambda'=\sqrt{N}$.   The second equation (\ref{sah}) can be rewritten as
 \beq
 \label{sah1}
\frac{\sqrt{N}}{g}=h\pm \sqrt{\frac{1}{\lambda}}\sqrt{\sigma}
 \eeq
 Thus, comparing to the $\sigma$-model gap equation (\ref{gapON2}) we see that we should keep the minus sign in (\ref{sah1}) and further interpret 
 \beq
 \label{lh}
 \lambda = \frac{16\pi^2}{N}\,,\,\,\,\,
 h=\frac{\sqrt{N}}{g_*}\,.
 \eeq
 The introduction of both $\Phi$ and $\Sigma$ clearly breaks higher spin symmetry. However, we expect that it is recovered only at the critical points. The free UV fixed point is reached taking $g,\lambda\to0$ and the cutoff to infinity, whereby $\sigma$ decouples. Therefore  only the $\Delta=1$ operator survives at the UV fixed point.
 
 On the other hand, the nontrivial IR fixed point arises when  $g\to g_*$. 
 Inspection of (\ref{Gas}) shows that the introduction of the operator $\alpha$ is equivalent to a finite shift of the operator $\sigma$. Hence, at $g=g_*$ the operator $\alpha$ becomes {\it redundant} and decouples from the theory.  

The $(\alpha-h)^3$ term  has an interpretation in terms of $(\Phi^a\Phi^a)^3$,  the (classically) marginal term, while $h$ introduces relevant terms in order that the non-trivial fixed point is properly described and appears at a finite value of $g$. This is equivalent to the well-known property that  {\it any} relevant deformation of the UV free fixed point will lead to the nontrivial IR theory \cite{Zinn_book}.

 
 \subsection{The singleton deformation of higher-spin theory and boundary symmetry breaking}

 Next, we deform the higher-spin action (\ref{HSaction}) by a singleton field $S$ of $m^2L^2=-\frac54$ as
 \beq
 \label{HSactiondef}
 I_{dHS}=I_{extHS}+\int d^4x\sqrt{-g}\frac{1}{2}S\left[\Box+\frac{5}{4L^2}\right]S\,,
 \eeq
 The asymptotic behaviour of $S$ is 
  \beq
 S\sim \xi z^{1/2}+\phi z^{5/2}.
\eeq
For such  a field, the only (unitary) possibility is to do AQ \cite{Marolf}, giving an operator of $\Delta=1/2$, a free field. Such a field must decouple from the rest of the CFT. However, it can be forced to have a non-trivial effect by coupling it to the other fields through an explicit boundary interaction,
namely $f(\phi,\alpha,\sigma)=\tilde\lambda\sigma\phi^2$. That this interaction is needed could have been anticipated from our discussion below eq. (\ref{Seff}) where we noted the presence of the $\sigma\varphi^2$ term as being crucial for the symmetry breaking structure of the theory.

Explicitly, we add to the deformed action (\ref{HSactiondef}) the following boundary term
\beq
\label{HSbdydeform}
f_d(\alpha,\sigma,\phi) = \int \left[\alpha\sigma -\tilde{V}(\sigma)-\lambda\frac{1}{3}\left(\alpha-h\right)^3+\tilde{\lambda}\sigma\phi^2 \right]\,,\,\,\,\tilde{V}(\sigma)=\frac{\tilde\lambda'}{g}\sigma\,,
\eeq
where using the results of the previous section, $h=\frac{\sqrt{N}}{g_*}$ and $\lambda=\frac{16\pi^2}{N}$. Other than the presence of the marginal term, a crucial difference between (\ref{HSbdydeform}) and (\ref{fas}) is in the linear deformation $\tilde{V}(\sigma)$, in which we have modified $\lambda'$ to $\tilde\lambda'=\frac{N+1}{\sqrt{N}}$. As discussed in Sec. 2.2, this is designed such as to be able to absorb the singleton field $\phi$ by suitably adjusting the coupling $1/g$ in the massive phase of the theory. The gap equations following from (\ref{HSbdydeform}) are
\beqn
\label{ag1}
\alpha+\tilde{\lambda} \phi^2&=&\frac{N+1}{\sqrt{N}}\frac{1}{g}\\
\label{sah2}
\sigma&=&\frac{16\pi^2}{N} \left(\alpha-\frac{\sqrt{N}}{g_*}\right)^2\\
\label{sphi}
\tilde{\lambda}\phi\sigma &=& 0
\eeqn
The third equation is familiar from the $\sigma$-model: there are two phases, one in which $\phi=0$ (massive phase) and the other in which $\sigma=0$ (broken phase). The first equation has an $O(N+1)$-invariant form if we interpret $\alpha\sim\langle \Phi^a\Phi^a\rangle$ and $\phi\sim\langle \Phi^{N+1}\rangle$. 
Substituting then $\alpha$ from (\ref{sah2}) to (\ref{ag1}) we find
 \beq
\tilde{\lambda} \phi^2=\frac{N+1}{\sqrt{N}}\frac{1}{g}-\frac{\sqrt{N}}{g_*}+\frac{\sqrt{N}}{4\pi^2}\sqrt{\sigma}\,.
 \eeq
Setting $\tilde{\lambda}=1/\sqrt{N}$ this coincides exactly with  (\ref{gap1}). The two solutions are
\beqn
1: \ \ \phi=0,\ \ \ \ \alpha=\frac{N+1}{\sqrt{N}}\frac{1}{g},\ \ \ \ \sigma=16\pi^2\left(\frac{N+1}{N}\frac{1}{g}-\frac{1}{g_*}\right)^2\\
2:\ \ \sigma=0,\ \ \ \ \ \alpha=\frac{\sqrt{N}}{g_*},\ \ \ \ \frac{1}{N}\phi^2=\left(\frac{N+1}{N}\frac{1}{g}-\frac{1}{g_*}\right)
\eeqn
Note the important point that $\alpha\neq 0$ does {\it not} signal $O(N)$ breaking.\footnote{That is, $\alpha$ is properly interpreted as the vev of an $O(N)$-invariant operator.} Rather $\phi\neq0$ implies $O(N+1)\to O(N)$. As before, there is a critical point when $g/g_*=(N+1)/N$. We can have $O(N+1)$ breaking only when $g/g_*<(N+1)/N$. For $g/g_*>(N+1)/N$, the only solution to the gap equations is of the first type, namely the massive phase. 




\subsection{The calculation of boundary anomalous dimensions}
At the critical point the operator $\alpha$ becomes redundant and the boundary term (\ref{HSbdydeform}) becomes 
\beq
\label{HSbdydeform1}
f_d(\sigma,\phi^2)=\frac{1}{\sqrt{N}}\int\sigma\phi^2\,.
\eeq
This is a rather simple marginal deformation of the extended higher-spin action (\ref{HSactiondef}) and leads to a $1/N$ expansion for the boundary two-point functions of $\phi$ and $\sigma$. For example, we obtain
\beq
\label{ppdeform}
\langle\phi(x_1)\phi(x_2)\rangle_{def} = \langle\phi(x_1)\phi(x_2)\rangle_0+\frac{1}{2N}\int d^3x \,d^3y\,\langle\phi(x_1)\phi(x_2)\sigma(x)\phi^2(x)\sigma(y)\phi^2(y)\rangle_0+\cdots\,,
\eeq
where we have dropped the $O(1/\sqrt{N})$ term whose contribution vanishes, as do all other fractional powers of $1/N$. 
Eq. (\ref{ppdeform}) gives {\it the same} expansion as in the field theory analysis, at least to leading order in $1/N$. 
Hence, the deformation (\ref{HSbdydeform1}) gives  for the boundary singleton field $\phi$ {\it the same} anomalous dimension as those  for the  UV dimensions of the elementary fields $\phi^a$, even though (\ref{HSbdydeform1}) may be regarded as a marginal deformation of the IR $O(N)$ fixed point in the presence of an additional scalar $\phi$.  Quite generally is not hard to see that the boundary graphical expansion for $\phi$ and $\sigma$ generated by (\ref{HSbdydeform1}) is the same as the graphical expansion for $\phi^a$ and $\sigma$ generated by the field theory (\ref{Zphisigma}), and hence yields the same anomalous dimensions. Ultimately, this works at leading order in $1/N$ precisely because one expects that any bulk diagrams do not lead to anomalous dimensions as long as higher spin symmetry pertains.

\section{Summary and outlook}

A complete holographic description of the $O(N)$ vector model should account for its rich vacuum structure and in particular for its $O(N)\rightarrow O(N-1)$ symmetry breaking pattern. In this work we have shown that this is possible if one deforms the AdS$_4$ higher-spin theory by a singleton field which sees the higher-spin multiplet only through a boundary marginal coupling. Then, imposing boundary conditions by the appropriate boundary terms to the extended bulk action we were able to exactly reproduce the gap equations of the $O(N)$ vector model. In doing so, we have discovered that the bulk higher-spin theory absorbs the singleton field by shifting its parameter $N\rightarrow N+1$. This is the bulk dual of the global symmetry breaking/enhancement mechanism in the boundary. Moreover, the boundary  interaction of the singleton with the higher-spin multiplet generates the same $1/N$ graphical expansion for the  elementary scalar and "spin-zero current" as in the standard field theoretic treatment of the $O(N)$ model. Hence, the singleton deformation breaks higher-spin symmetry and yields the well-known anomalous dimensions for the elementary and "spin-zero" scalars of the $O(N)$ model, at least to leading order in $1/N$. 

There is a number of immediate extensions to our work. Firstly, is it important to understand better the boundary marginal coupling of the singleton to higher-spin currents. For example,
given the singleton field $\phi$, one may consider boundary couplings of the form 
\beq
\label{HScouplings}
S_{HS}\sim \lambda'\int t^{\mu_1...\mu_s}\phi\pa_{\mu_1}...\pa_{\mu_s}\phi\,,
\eeq
where $t^{\mu_1..\mu_s}$ is the {\it leading} coefficient in the asymptotic behaviour of a bulk spin-$s$ gauge field. For $s\geq 2$ there are more than one possible term in (\ref{HScouplings}).   Once again, we expect that $\lambda'\sim 1/\sqrt{N}$. Generally, this has no effect on the vacuum structure, if that is determined by space-time constant configurations. It is  expected that (\ref{HScouplings}) would lead to a graphical expansion for the 2-pt functions of the boundary higher-spin currents which would enable one to calculate their $1/N$ anomalous dimensions. Reproducing the result (\ref{HSanomdims}) would then be a crucial test for our proposal. 

Our results can also be applied to the holographic description of three-dimensional fermionic and supersymmetric models with higher-spin duals \cite{LP1,SS1}. Notice that such models describe parity symmetry breaking, and it would be interesting to understand the bulk counterpart of it.

In AdS$_5$/CFT$_4$ correspondence adding a probe D3-brane to IIB sugra on AdS$_5\times S^5$ shifts by one unit $N\rightarrow N+1$ the fiveform flux. The singleton deformation is the analog process of the above in higher-spin gauge theory and its study might lead to a better geometric description for the dimensionless parameter $N$.  The singleton deformation could also play an important role in the study of possible black-hole solutions for higher-spin theory on AdS$_4$. For example, since a continuous symmetry cannot be broken at finite temperature in 2+1 dimensions, we expect that bosonic singleton absorption would not be possible for higher-spin theories in black-hole backgrounds, while it should be possible for fermionic singletons.   We hope to report on these issues in the near future \cite{LPnew}.

\acknowledgments
ACP would like to thank CPHT in \'Ecole Polytechnique, Paris, where he spent three months as an invited CNRS Researcher, for its hospitality and excellent working environment provided during the  completion of this work. We would like to thank C. Bachas, D. Minic, L. Pando Zayas, P. M. Petropoulos, D. Vaman and A. Zaffaroni for helpful discussions. The work of ACP was partially supported by the research program ``AdS/CMT - Holography and Condensed Matter Physics" (ERC - 05), MIS 37407, by the Greek government. It was also cofinanced by the European Union (European Social Fund, ESF) and Greek national funds
through the Operational Program Education and Lifelong Learning" of the National Strategic Reference
Framework (NSRF) under Funding of proposals that have received a positive evaluation in the 3rd and
4th Call of ERC Grant Schemes. RGL is supported in part by U.S. Dept. of Energy grant FG02-91-ER40709 and is grateful to the Perimeter Institute for Theoretical Physics for support through their sabbatical program.

\appendix

\begin{appendix}

\section{Propagators and vertices}
The propagators and vertices we use in the text are given by
\bea
\label{Delta}
\Delta(x,y;\rho_0) &=& -\frac{1}{2}\int\frac{d^3 p}{(2\pi)^3}e^{ip(x-y)}\Pi(p;\rho_0)\,,\\
\Pi(p;\rho_0)&=&\int\frac{d^3q}{(2\pi)^3}\frac{1}{(q^2+\rho_0)[(p+q)^2+\rho_0]}\,\\
P(x,y,z;\rho_0)&=& D_0^{-1}(x,y;\rho_0)D_{0}^{-1}(y,z;\rho_0)D_{0}^{-1}(z,x;\rho_0)\,,\\
D_0^{-1}(x,y;\rho_0) &=& \int\frac{d^3p}{(2\pi)^3}\frac{e^{ip(x-y)}}{p^2+\rho_0}\,.
\eea

To leading order in $1/N$ the two-point functions for the fields $\phi^a$ and $\sigma$ at the the nontrivial critical point $g=g_*$ are 
\bea
\label{phiphi}
\langle\phi^a(x)\phi^b(y)\rangle &\equiv& D_0^{-1}(x,y;0)\delta^{ab}=\frac{1}{4\pi}\frac{1}{|x-y|}\delta^{ab}\\
\label{ll}
\langle\sigma(x)\sigma(y)\rangle &\equiv& 
\Delta^{-1}(x,y;0)=\frac{16}{\pi^2}\frac{1}{|x-y|^4}\,,
\eea
Notice that the momentum space 2-pt function (\ref{Delta}) is negative since the Lagrange multiplier field $\sigma$ is imaginary \cite{Zinn_book}.  

The massless diagrams are evaluated using 
\bea
\label{integral}
\int d^d x\frac{1}{|x-x_1|^{2A}|x-x_2|^{2B}}&=&C(A,B,d)\frac{1}{|x_1-x_2|^{2(A+B-\frac{d}{2})}}\\
\label{CABd}
C(A,B,d) &=&\pi^{\frac{d}{2}}\frac{\Gamma(\frac{d}{2}-A)\Gamma(\frac{d}{2}-B)\Gamma(A+B-\frac{d}{2})}{\Gamma(A)\Gamma(B)\Gamma(d-A-B)}
\eea





\section{Functional integrals and vacuum structure of field theories}
We review here some standard results (e.g. see \cite{Zinn_book}) regarding the study of vacuum structure in field theories using functional integrals. If $S$ is the action functional of the theory, the Euclidean generating functional $W[J]$ for connected correlation functions of the composite operator ${\cal O}$ is defined as
\beq
\label{genfun}
Z[J]=\int e^{-S+\int J{\cal O}} = e^{W[J]}\,,
\eeq
where $J$ is the source. The quantum effective action $\Gamma[\sigma]$ (generating functional for 1PI graphs) is the Legendre transform of $W[J]$ and we have the usual relations
\beq
\label{1PI}
W[J]+\Gamma[\sigma]=\int J\sigma\,,\,\,\,\frac{\delta W[J]}{\delta J}=\langle{\cal O}\rangle_J\equiv \sigma\,,\,\,\,\frac{\delta\Gamma[\sigma]}{\delta\sigma}=J\,.
\eeq
The effective potential $U_{eff}$ (vacuum energy density) is  given by 
\beq
\label{effpot}
\Gamma[\sigma_{ct}]=\frac{1}{Vol}U_{eff}[\sigma_{ct}]\,,
\eeq
where $\sigma_{ct}$ are constant configurations. Setting the source $J$ to zero yields the gap equation 
\beq
\label{vacuum}
\frac{\delta \Gamma[\sigma]}{\delta\sigma}\Bigl|_{\sigma_{*}}=0\,,
\eeq
which determines the vacuum structure of the theory, namely the values $\sigma_*$ for which the vacuum energy  is extremized.

Suppose now that we want to calculate the generating functional of ${\cal O}$ in a deformed theory, namely
\beq
\label{Sdeform1}
e^{W_f[J]}=\int e^{-S_f+\int J{\cal O}}\,,\,\,\,S_f=S+\int f({\cal O})
\eeq
Assuming that the theory has a large-$N$ expansion we find
\beq
\label{WJf}
e^{W_f[J]}\approx \int e^{-S+\int J{\cal O}-\int f(\langle{\cal O}\rangle)}=e^{W[J]-\int f(\langle{\cal O}\rangle)}
\eeq
Then, using (\ref{1PI}) we can find the effective action of the deformed theory as
\beq
\label{Gammaf}
\Gamma_f[\sigma]\approx\Gamma[\sigma]+\int f(\sigma)
\eeq

\subsection{Vacuum structure from holography}

Consider the bulk action $I_b$ for a scalar field\footnote{We use the usual Poincar\'e  patch for AdS$_4$. $z$ denotes the radial coordinate, $x$ the boundary spacetime coordinates. The boundary is at $z\rightarrow 0$.}  $\phi(z,x)$ on AdS$_4$ which we take here, without loss of generality and for later convenience, to be a conformally coupled scalar.
The boundary behaviour of $\phi$ is 
\beq
\label{scalar_asympt}
\phi(z,x)\sim z\alpha(x) +z^2\beta(x)\,.
\eeq
Then, the on-shell variation of the renormalized bulk action is, in the standard quantization,
\beq
\label{onshellvar}
\delta I_R^{o.s.}=\delta(I+I_{c.t.})^{o.s.} =-\int \beta\delta\alpha
\eeq
where $I_{c.t.}$ are the appropriate counterterms (including generalized Gibbons-Hawking terms). Formula (\ref{onshellvar}) implies that the bulk action is stationary for Dirichlet boundary conditions $\delta\alpha=0$. However, it also implies that we can identify the on-shell renormalized bulk action with the generating functional for connected correlation functions of a scalar operator ${\cal O}$ as
\beq
\label{osaction}
I^{o.s.}_{R}[\alpha=J]\equiv -W[J]\,,\,\,\,\,\frac{\delta W[J]}{\delta J}=\langle{\cal O}\rangle_J\equiv \beta\,.
\eeq
Hence, from the definition (\ref{genfun}) of $W[J]$ we learn that imposing the boundary condition $\alpha=J$ to the on-shell bulk action is equivalent to evaluating the boundary path integral for the linearly deformed action
\beq
\label{lindeform}
S\rightarrow S-\int J{\cal O}\,.
\eeq
For a bulk conformal scalar with $m^2L^2=-2$, ${\cal O}$ is an operator with dimension $\Delta=2$. From $W[J]$ we can find using (\ref{1PI}) the effective action $\Gamma[\beta]$ and then setting $J=0$ the gap equation of the boundary theory.

Consider now the deformed boundary action $S_f$ 
\beq
\label{Sdeform}
S_f=S+\int f({\cal O})\,,
\eeq
where $f({\cal O})$ is a local functional of the field ${\cal O}$. This is referred to as Lagrangian deformation. The  generating functional $W_f[J]$ of the deformed theory is  given by
\beq
\label{Wdeform}
e^{W_f[J]}=\int e^{-S_f+\int J{\cal O}}\,,
\eeq
which as  shown in Appendix B.1 lead to (\ref{Gammaf}). 
%
Functional differentiation that with respect to $\sigma$ yields  the gap equation of the deformed theory 
\beq
\label{gapdeformed}
\frac{\delta\Gamma_f}{\delta\sigma}\Big|_{\sigma=\sigma_*}=J+f'(\sigma_*)=0\,.
\eeq
The result (\ref{gapdeformed}), first derived in \cite{Witten}, can be used to describe holographically the vacuum structure of generic boundary theories. 

It is well-known that for a conformally coupled bulk scalar, there is an alternative quantization for which 
\beq
\delta I^{o.s.}_{AQ}=+\int \alpha\delta\beta
\eeq
In this case, $\beta$ plays the role of source for a $\Delta=1$ operator and we write
\beq
\label{osactionAQ}
I^{o.s.}_{AQ}[\beta=J]\equiv -W[J]\,,\,\,\,\,\frac{\delta W[J]}{\delta J}=\langle{\cal O}\rangle_J\equiv \alpha\,.
\eeq
Lagrangian deformation of this theory leads to similar equations as above.
The  above result is generic and in particular applies when the theory is strongly coupled. 



\end{appendix}






\end{document}